\RequirePackage{lineno} 
\documentclass[aps,prl,twocolumn]{revtex4}
\usepackage{amsmath}
\usepackage{amssymb}
\usepackage{graphicx}
\usepackage{color}
\usepackage{physics}
\usepackage{gensymb}
\usepackage{todonotes}
\usepackage{float} %For fig. positioning
\usepackage{xcolor} %For fig. positioning
\usepackage{cancel}%Antonis. During the drafts, to keep track of previous phrasings.

\pdfoutput=1

\newcommand{\eq}{\begin{equation}}
\newcommand{\eeq}{\end{equation}}

\begin{document}

%CM: trying to broaden the title
\title{Generation of Thermofield Double States and \\ Critical Ground States with a Quantum Computer}

%\title{Variational Quantum Simulation of Thermofield Double States with Trapped Ions/Variational Simulation of Quantum Systems at Finite-Temperature with Trapped Ions}

\author {D. Zhu$^{1*}$, S. Johri$^{2}$, N. M. Linke$^{1}$, K. A. Landsman$^{1}$,  N. H. Nguyen$^{1}$, C. H. Alderete$^{1,3}$, A. Y. Matsuura$^{2}$, T. H. Hsieh$^{4}$, and C. Monroe$^{1}$}

\affiliation{$^1$ Joint Quantum Institute, Center for Quantum Information and Computer Science, and Department of Physics, University of Maryland, College Park, Maryland 20742  USA}
\affiliation{$^2$ Intel Labs, Intel corporation, Hillsboro, Oregon 97124  USA}
\affiliation{$^3$ Instituto Nacional de Astrof\'{i}sica, \'{O}ptica y Electr\'{o}nica, Sta. Ma. Tonantzintla, Puebla
72840, Mexico
}
\affiliation{$^4$ Perimeter Institute for Theoretical Physics, Waterloo, Ontario N2L 2Y5, Canada}

\date{\today}

\begin{abstract}
Finite-temperature phases of many-body quantum systems are fundamental to phenomena ranging from condensed-matter physics to cosmology, yet they are generally difficult to simulate \cite{feynman1982}. Using an ion trap quantum computer and protocols motivated by the Quantum Approximate Optimization Algorithm (QAOA) \cite{farhi2014quantum}, we generate nontrivial thermal quantum states of the transverse-field Ising model (TFIM) by preparing thermofield double states \cite{TFDoriginal} at a variety of temperatures. We also prepare the critical state of the TFIM at zero temperature using quantum-classical hybrid optimization. The entanglement structure of thermofield double and critical states plays a key role in the study of black holes, and our work simulates such nontrivial structures on a quantum computer.  Moreover, we find that the variational quantum circuits exhibit noise thresholds above which the lowest depth QAOA circuits provide the best results.
%, contrary to theoretical expectations. [is this so surprising?]
%The finite-temperature phase diagram of quantum systems exhibits rich physics, but so far was not studied on the nascent quantum computers increasingly available today. Here we use a 7 qubit trapped-ion system to implement a variational state-preparation ansatz based on the bang-bang protocol of the Quantum Approximate Optimization Algorithm to prepare thermofield double states of the transverse field Ising model (TFIM) at a variety of temperatures. Thermofield double states are important in the study of gravitational physics, and also serve as purifications of the Gibbs states of the TFIM. We also implement a QAOA-like technique to directly prepare the critical state of the TFIM at $T=0$. Our $T=0$ protocol shows that the lowest depth ansatz gives the best results, contrary to theoretical expectations. Our simulation of noise in this experiment shows a threshold-like behavior in the response of these variational quantum circuits to noise.
\end{abstract}


\maketitle %\maketitle must follow title, authors, abstract and \pacs


%\section{Introduction}
Progress in the control of synthetic quantum systems such as superconducting qubits \cite{devoret13} and trapped ions \cite{monroe13} has enabled continual advances in the depth of quantum computer circuits and the complexity of quantum simulations.  As the number of qubits and their coherence times increase, such systems have the potential to simulate highly non-trivial macroscopic quantum phenomena. 
%There is a wealth of interesting many-body quantum states that cannot be directly studied have yet to be realized in solid-state materials, and synthetic quantum systems provide an alternative venue to systematically construct such states from scratch.  
While there has been progress in the preparation of entangled quantum states such as squeezed or ``cat'' states \cite{Haroche2013, Kasevich2016}, much less attention has been paid to generating thermal (Gibbs) states of a many-body Hamiltonian, even though these states underpin phenomena ranging from high temperature superconductivity \cite{lee2006doping} to quark confinement in quantum chromodynamics \cite{gross}.
%topological states of matter \cite{Haldane2017}.
%While quantum phases of matter and their transitions are typically considered at zero temperature, many finite temperature phenomena of quantum systems eludes theoretical understanding.  For example, a typical phase diagram for high temperature superconductors \cite{lee2006doping} is filled with exotic labels such as ``pseudo-gaps'' and ``strange metals'' that beg for more experimental and theoretical insight. 

The simulation of many-body thermal states challenges currently available quantum platforms, owing to the required level of control over both the many-body interactions and the effective coupling to the thermal bath.  Proposed schemes \cite{temme2011quantum, terhal, poulin} to generate many-body thermal states involve subroutines like quantum phase estimation, which are difficult to implement on near-term devices, or require engineered dissipative couplings \cite{brandao}. Experimental platforms such as optical lattices of ultracold atoms have enabled finite temperature simulation \cite{jordens2008mott,gross2017quantum}, but these are specific to particular (Hubbard) models, and cooling to low effective temperatures remains a major obstacle. 

%Here we use a very different approach for simulating thermal Gibbs states by preparing 
Here we use an ion trap quantum computer to generate various nontrivial quantum states in the context of the many-body transverse field Ising model. We generate thermofield double (TFD) states \cite{TFDoriginal}, which are pure quantum states entangled between two systems, with the property that when either system is considered independently by tracing over the other, the TFD reduces to a thermal mixed state at a specified temperature.
%thermal analogues of pure EPR/Bell entangled states.  
TFD states are purifications of thermal Gibbs states and have played a key role in the holographic correspondence relating a quantum field theory to a gravitational theory in one higher dimension.  In this correspondence, TFD states are dual to wormholes on the gravity side \cite{TFD1, TFD2} and enable teleportation (``traversable wormholes'') \cite{traversable1, traversable2}. The simulation of these concepts has motivated several approaches for preparing TFD states \cite{wu2018variational, martynswingle, lokhande, qimaldacena}.  

\begin{figure*}%[htbp]
\centering
\includegraphics[width=1.7\columnwidth]{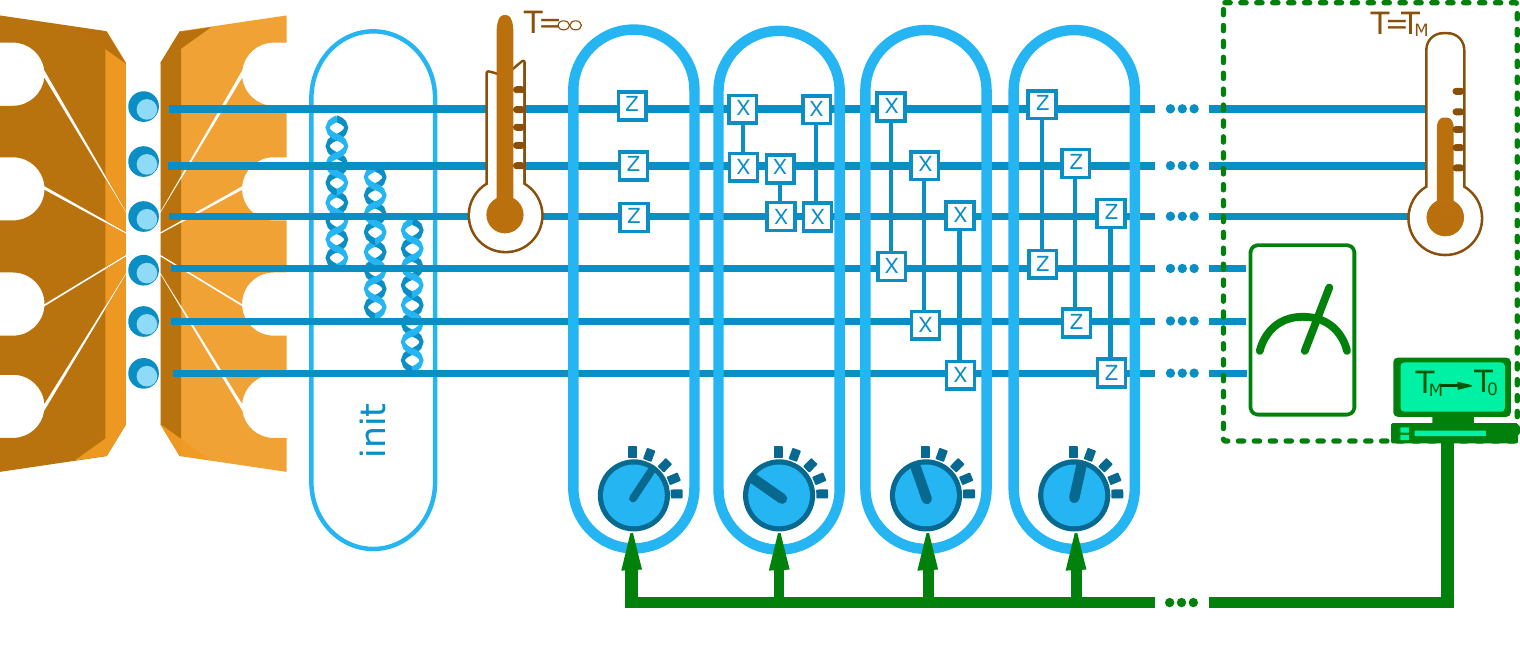}
 \caption{Hybrid quantum-classical optimization with trapped ion qubits to prepare thermal states. The initial Bell-pairs correspond to the thermofield double state at infinite temperature. Layers of unitaries with independent control parameters are then applied sequentially to cool to the target temperature. The subsystem consisting of the first three qubits is effectively in the thermal (Gibbs) state. The result can be fed into a classical computer which updates the parameters based on a cost-function in a closed loop (see ``Full Hybrid Optimization: Preparation of Ground State of TFIM" for details).} \label{fig:QAOA_Loop}
 \end{figure*}

In this work, we use protocols \cite{wu2018variational} inspired by the alternation of unitary operators that forms the basis of the quantum approximate optimization algorithm (QAOA) \cite{farhi2014quantum}. This scheme allows us to use unitary operations to control the effective temperature of a subsystem, thus foregoing the need of an external heat bath.  We prepare TFD states of the quantum critical transverse field Ising model in a ring geometry composed of three trapped ion effective spins, at various target temperatures, as shown in Fig. \ref{fig:QAOA_Loop}. We also use a related approach \cite{ho2018efficient} to directly prepare the zero temperature ground state of the quantum critical transverse field Ising model with seven trapped ion spins using quantum-classical feedback. 

% MOVE BELOW PARAGRAPH TO SUMMARY?
%Our work demonstrates the versatility of both variational quantum simulation of the QAOA and the trapped ion platform in preparing a variety of pure and thermal quantum states of interest. Our TFD state preparation paves the way for finite temperature quantum simulation without an external heat bath. Furthermore, our analysis of the experiments also provides insight into how variational, QAOA-type protocols are impacted by experimental errors. 

\section{Thermofield Double States}

We briefly review the definition and preparation scheme of the TFD state.  Consider two identical Hilbert spaces $A$ and $B$ consisting of qubits labeled by an index $j$. Let $H_A$ be a Hamiltonian with eigenstates $\ket{n}_A$ and corresponding energies $E_n$.  A thermofield double state corresponding to inverse temperature $\beta$ is defined on the joint system A and B as

\begin{equation}\label{eq:TFD}
    \ket{TFD(\beta)}=\frac{1}{Z(\beta)}\sum_{n}e^{-\beta E_n/2}\ket{n}_A\ket{n'}_B
\end{equation}
where $Z(\beta)$ is a normalization factor.  In general, the set $\{\ket{n'}_B\}$ can be any orthonormal basis spanning $B$, and we will make the choice $\ket{n'}=\Theta \ket{n}$, where $\Theta$ is the time-reversal operator $\otimes_j iY_j K$, where $K$ is complex conjugation and $Y_j$ is the Pauli-Y operator at site $j$.  This choice is consistent with the infinite temperature TFD defined below. Tracing out the auxiliary system $B$ results in the thermal (Gibbs) state of system A $\rho_A = e^{-\beta H_A}/Z$; in this sense, realizing the TFD allows one to simulate the thermal Gibbs state in a subsystem $A$ with the effective bath $B$.

The protocol \cite{wu2018variational} starts with an initial state $\ket{\psi_0}$ that is a product of Bell-pair singlets $\frac{1}{\sqrt{2}}(\ket{0}\ket{1}-\ket{1}\ket{0})$ between pairs of $A$ and $B$ qubits. This is an infinite temperature TFD since $\rho_A$ is maximally mixed. Note that the two components of a Bell-pair singlet are related by time-reversal symmetry ($\Theta \ket{0}=\ket{1}$ and $\Theta \ket{1}= -\ket{0}$), which justifies our choice of basis above. One then alternates between time evolution with the inter-system coupling $H_{AB}=\sum_i X_{i,A} X_{i,B} + Z_{i,A} Z_{i,B}$ and the intra-system Hamiltonians $H_A+H_B$, where $H_B$ is the time-reversed version of $H_A$.  As in QAOA, each timestep is a variational parameter, and after $p$ layers of alternation, the resulting variational wavefunction is:  
\begin{eqnarray}
| \psi(\vec{\alpha},\vec{\gamma})\rangle_p = \prod_{j=1}^{p} e^{i\alpha_j H_{AB}} e^{i\gamma_j (H_A + H_B)/2} \ket{\psi_0} 
\label{eq:ansatz}
\end{eqnarray}
The variational parameters $\vec{\alpha},\vec{\gamma}$ are chosen to maximize the fidelity with the target TFD state: $F_p(\vec{\alpha},\vec{\gamma}) \equiv |\langle \mathrm{TFD}(\beta)| \psi(\vec{\alpha},\vec{\gamma})\rangle_p|^2$.  As detailed in \cite{wu2018variational}, this protocol is guaranteed to target the zero temperature TFD in the limit of large $p$ because in that limit it subsumes the adiabatic algorithm; the intuition, verified through several examples \cite{wu2018variational}, is that the finite temperature TFD is easier to prepare than zero temperature ground state because the thermal correlation length is generally finite.

%\section{Thermofield Double State of a Quantum Critical Point}

 In the holographic correspondence, TFDs of conformal field theories describing gapless quantum matter are particularly interesting because they correspond to wormholes on the gravity side.  Their preparation is also useful to condensed matter physics because they enable investigation of finite-temperature properties of systems near a critical point by tracing over one of the systems in the double. Hence, our first objective is to prepare thermofield double states of the transverse field Ising model (TFIM) at its quantum critical point. Defined on a one-dimensional ring of $L$ qubits, the TFIM Hamiltonian is 

\begin{equation}\label{eq:H_TFIM}
    H_{TFIM}=\sum_{i=1}^{L}X_iX_{i+1} + g \sum_{i=1}^{L}Z_i \equiv H_{XX} + g H_Z
\end{equation}
Here $g$ is the strength of the transverse field.  When $g=1$, the ground state is a critical point between anti-ferromagnetic and paramagnetic quantum phases and has several interesting properties, including correlations between two spins decaying as a power of their separation and entanglement entropy scaling logarithmically with the size of the subsystem.  

To prepare the TFD of the quantum critical TFIM, we tailor the general protocol above (Eq. \ref{eq:ansatz}) to the capabilities of an experimental system with six trapped ions.  The initial state is the product state of three spin-singlet Bell pairs formed between pairs of A and B spins. Ideally following the general protocol, we would like to evolve sequentially with $H_A=H_{XX}+H_Z$ (in addition to the time-reversed copy of the Hamiltonian on the $B$ system $H_B$), followed by

\begin{equation}\label{eq:H_AB_TFIM}
  H_{AB}=\sum_{i}Z_{i,A}Z_{i,B}+ \sum_{i}X_{i,A}X_{i,B} \equiv H_{ABZ}+H_{ABX}.
\end{equation}
Since $H_{ABZ},H_{ABX}$ commute, this step can be simply decomposed into evolution with each piece separately.  
However, time evolution with $H_A$ in general requires a Trotter decomposition which could require many steps beyond the capabilities of current experimental systems.  Moreover, here $H_B$ introduces additional gates which we find are not essential for achieving high fidelity.  Hence, we instead use a minimal variational ansatz for the TFD consisting of four pieces:

\begin{widetext}

\begin{equation}\label{eq:QAOA_process}
  \ket{\psi(\alpha_1, \alpha_2, \gamma_1, \gamma_2)}=\exp(iH_{ABZ}\alpha_2)\exp(iH_{ABX}\alpha_1)\exp(iH_{XX}\gamma_2)\exp(iH_Z\gamma_1)\ket{\psi_0}
\end{equation}

\end{widetext}

\begin{figure*}[htbp]
\centering
\includegraphics[width=1.5\columnwidth]{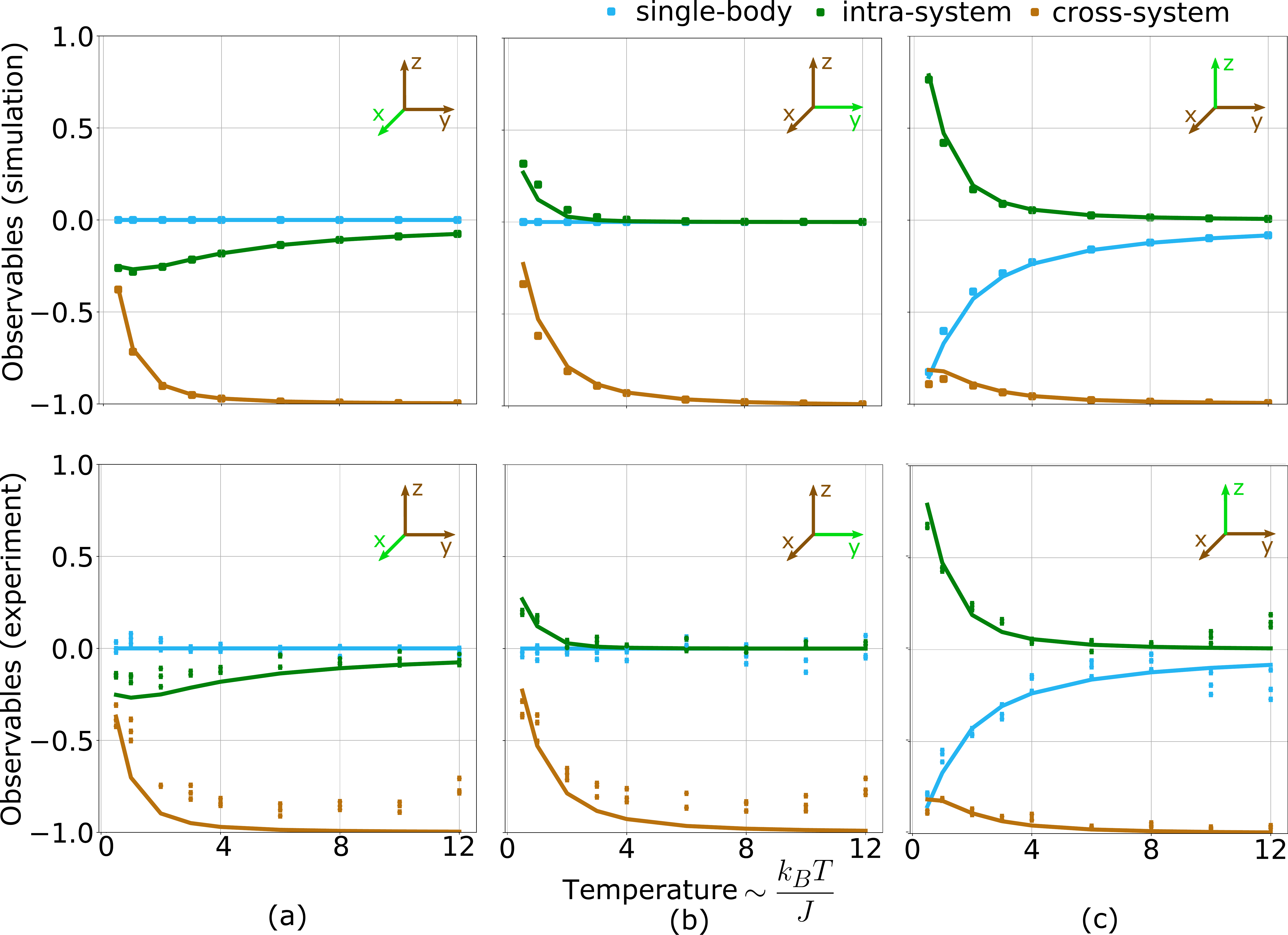}
 \caption{Preparation of TFD states of the quantum critical TFIM  using two 3-qubit systems. Top row: Comparison between observables of the simulated optimized ansatz circuit and target TFD states (dotted line) for various target temperatures. Bottom row: Comparison between observables of experimentally prepared and target TFD states. Results for all three ion pairs are given at each temperature. The measured correlation functions for different target temperatures are plotted against the theoretical expectations (dotted lines) for type (a) Pauli-X  (b) Pauli-Y and (c) Pauli-Z. Intra-system correlators in the subsystem-A are: $<\sigma_{1,A}\sigma_{2,A}>$, $<\sigma_{1,A}\sigma_{3,A}>$, and $<\sigma_{2,A}\sigma_{3,A}>$.Cross-system correlators are $<\sigma_{1,A}\sigma_{1,B}>$, $<\sigma_{2,A}\sigma_{2,B}>$, and $<\sigma_{3,A}\sigma_{3,B}>$. Note the experimental data points in the figure have errorbars accounting for statistical errors. Statistical error bars are similar in  size or smaller than the symbols used. A symmetry based error mitigation technique is used to post-process the experimental result in (c). The mitigation notably improved the agreement between experiment and theory. Details are given in the appendix.} \label{fig:TFD_TFIM}
 \end{figure*}

The first two operations represent a minimal Trotterization of time evolution with $H_A$.  The optimal parameters are determined (on a classical computer) by maximizing the fidelity with the target TFD.  In this case, the optimal fidelities are extremely good, ranging from $0.93$ for the zero temperature TFD to $1$ for the infinite temperature TFD.  These can be further improved by adding additional iterations of this sequence of unitaries in the protocol.  The single-body observables and two point correlation functions of the optimized ansatz compare well with those of the target TFD, as evident in Fig.\ref{fig:TFD_TFIM}.  We note that the general protocol preparing the TFD of the classical ($g=0$) Ising model achieves perfect fidelity for $p=L/2$ layers \cite{wu2018variational}. 

%\begin{equation}\label{eq:H_Z_TFIM}
 % H_{Z}=\sum_{i}Z_{i,A}
%\end{equation}

%\begin{equation}\label{eq:H_XX_TFIM}
 % H_{XX}=\sum_{i}X_{i,A}X_{i+1,A}
%\end{equation}

%\begin{equation}\label{eq:H_ABX_TFIM}
 % H_{ABX}=\sum_{i}X_{i,A}X_{i,B}
%\end{equation}

%\begin{equation}\label{eq:H_ABZ_TFIM}
%  H_{ABZ}=\sum_{i}Z_{i,A}Z_{i,B}
%\end{equation}

We experimentally run the optimized state-generation protocol on an ion trap quantum computer (see appendix for experimental details). To confirm the preparation of the TFD state, we measure both intra-system observables (single and two body correlation functions within system $A$) and inter-system correlators between corresponding sites from the $A$ and $B$ systems.  The purpose of the intra-system measurements is to verify physical properties of the thermal Gibbs state.  In the phase diagram parameterized by temperature $T$ and transverse field $g$, there is a regime $|g-1|<<T<<1$ called the quantum critical fan \cite{subir}, whose properties are dictated by the continuum theory of the critical point.  For instance, this regime exhibits exponentially decaying correlations with correlation length proportional to inverse temperature in this case.  Our intra-system measurements could verify this phenomena and other features of the quantum critical fan for larger system sizes.  The purpose of the inter-system measurements is to observe how correlations and entanglement between the two systems decreases as one lowers the target temperature and thereby the thermal entropy (which in the TFD is the entanglement entropy between the systems).  

As shown in Fig.\ref{fig:TFD_TFIM}, the results agree well with those expected from the TFD states, with some reduction in correlations caused by imperfect entangling operations.  We note that at high temperature, there is a slight increase in error arising from an artifact of the optimization landscape being nearly degenerate; there are many sets of parameters that yield very good fidelities, and the optimal angles found are large enough to cause the observed errors. In fact, for such high temperatures, the initial set of Bell pairs is already a very good approximation to the target TFD, and it would be better to avoid using any gates.   

\section{Quantum Critical State at $T=0$}

To prepare the zero-temperature critical TFIM (pure) state, one does not require a purifying auxiliary system and thus a larger system $A$ can be accessed experimentally.  However, the long-range correlations and relatively high entanglement of the critical state pose challenges for preparation. Because a finite depth circuit consisting of local gates can only produce a state with finite correlation length, to generate critical states one needs a quantum circuit (of local gates) with depth scaling with system size. With non-local gates, long range correlated states can be prepared with fewer steps \cite{PhysRevA.99.052332}; however, tailoring the effective power-law decaying interactions in trapped ion systems to target an arbitrary critical state is in general a difficult problem.  One method for generating such critical states is the adiabatic algorithm, which requires tuning $g$ adiabatically. On a digital quantum platform, this would require a compilation such as Trotterization into discrete gates, and the resulting deep circuit would be very susceptible to errors.  

An alternative is the QAOA-motivated variational approach detailed in \cite{ho2018efficient}.  One begins with the product ground state of $H_Z$, which we denote $|0\rangle$, and then evolves with $H_{XX}, H_Z$ in an alternating fashion: 
\begin{equation}\label{eq:QAOA}
\ket{\psi(\vec{\alpha},\vec{\gamma})}_p=e^{-i\alpha_pH_Z}e^{-i\gamma_pH_{XX}}\cdots e^{-i\alpha_1H_Z}e^{-i\gamma_1H_{XX}}\ket{0}
\end{equation}
Again, $p$ denotes how many pairs of iterations are used, and $(\vec{\gamma},\vec{\beta})$ are variational parameters to be optimized to minimize the energy cost function
\begin{equation}
E_p(\vec{\alpha},\vec{\gamma}) = {_p\langle} \psi(\vec{\alpha},\vec{\gamma})| -H_{XX} - H_Z |\psi(\vec{\alpha},\vec{\gamma})\rangle_p
\end{equation}
The lower the energy, the better this wavefunction can approximate the critical ground state of $-H_{XX} - H_Z$.  Note the minus signs in the Hamiltonian and cost function; in this section, we target the critical ground state of the ferromagnetic transverse field Ising model.

Trotterizing the adiabatic approach for preparing the critical state would lead to a unitary sequence of the above form, with $(\vec{\gamma},\vec{\beta})$ infinitesimal; this implies that for sufficiently large numbers of layers $p$, there is guaranteed to exist a set of parameters $(\vec{\gamma},\vec{\beta})$ for which the ansatz converges to the target state.  However, the key question is how well the above ansatz performs for finite $p$.  Remarkably, it has been observed that for a system size $L$, the above protocol can prepare the target critical state (and any state in the TFIM phase diagram) with perfect fidelity given $p = L/2$ layers \cite{ho2018efficient}. 

\begin{figure}[htbp]
\centering
\includegraphics[width=\columnwidth]{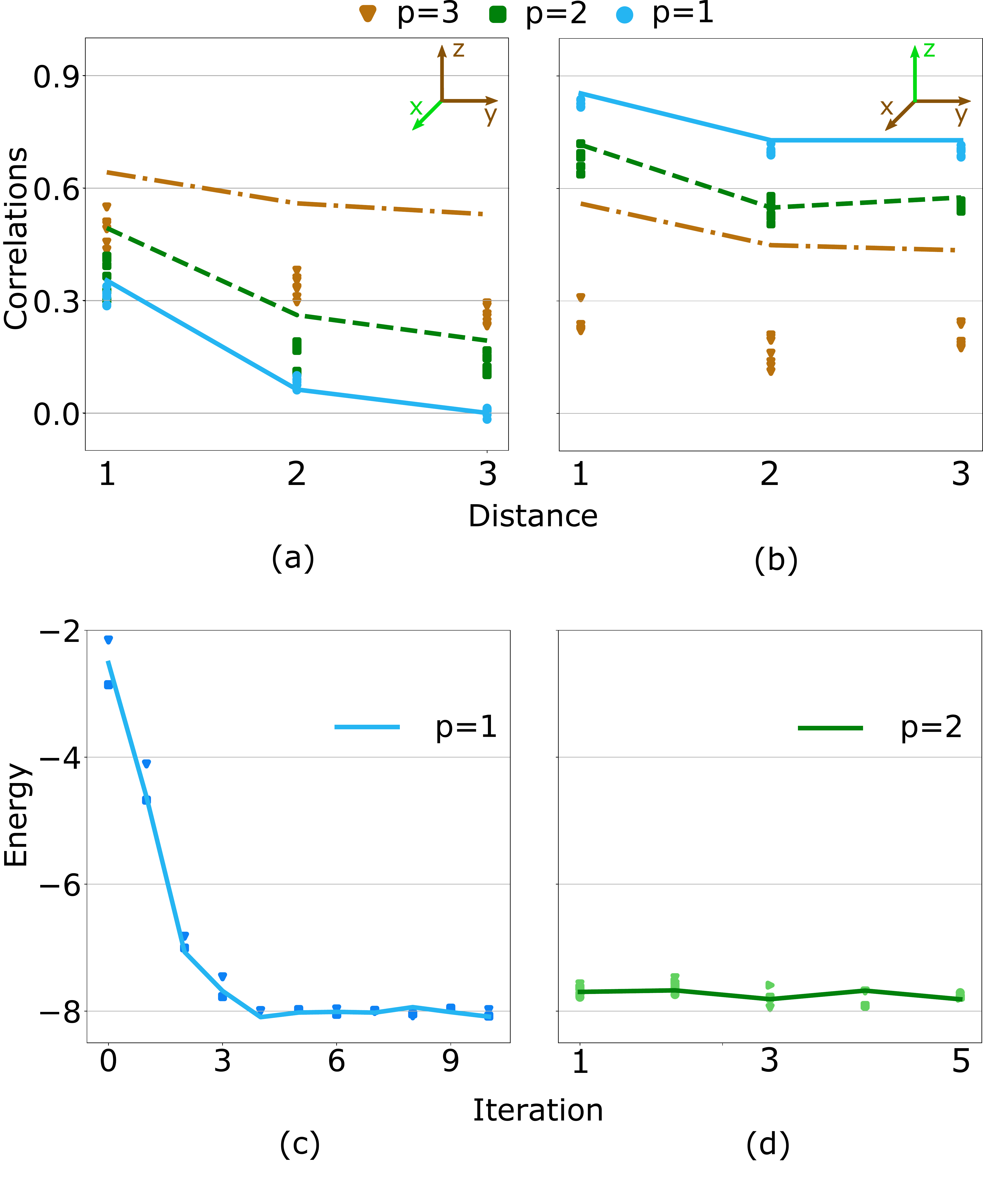}
\caption{Critical TFIM ground state on a 7-qubit system. Top row: Two-point correlations for (a) Pauli-X  and (b) Pauli-Z operators as a function of their separation. For a ring of seven spins, there are only three different pairs of ions, which are distinguishable by distance. The three different colors correspond to QAOA protocols with different depth $p$. The lines denote the theoretical expectations. Bottom row: Energies achieved using full hybrid quantum-classical feedback with increasing gradient descent iteration number for (c) $p=1$, initialized with random parameter set, and (d) $p=2$, initialized with theoretically optimal parameters. The line corresponds to the measured energy at each iteration, and the dots correspond to samples taken to evaluate the gradients. Ideally, the lowest energy a p=1 protocol can reach is $-8.44$. The lowest energy a p=2 protocol can reach is $-8.62$. The true ground state energy is $-8.98$, and the size of the gap is $0.23$. The gap decreases linearly with system size. Statisical error bars in the above figures are of  the same size or smaller than the symbols used.}  \label{fig:critical_state}
 \end{figure}

For a trapped ion system of seven qubits, a $p=3$ protocol can generate the desired ground state with perfect fidelity, and we find the optimal angles $(\vec{\alpha}, \vec{\gamma})$ on a classical computer to maximize the many-body overlap $\left|\bra{\psi_t}\ket{\psi_p}\right|^2$ of the ansatz $\ket{\psi_p}$ and the target state $\ket{\psi_t}$.  While $p=3$ layers exactly prepares the critical state, $p=1,2$ yield theoretical fidelities of 0.76 and 0.88, respectively.  

\begin{figure*}[htbp]
\centering
 \includegraphics[width=1.5\columnwidth]{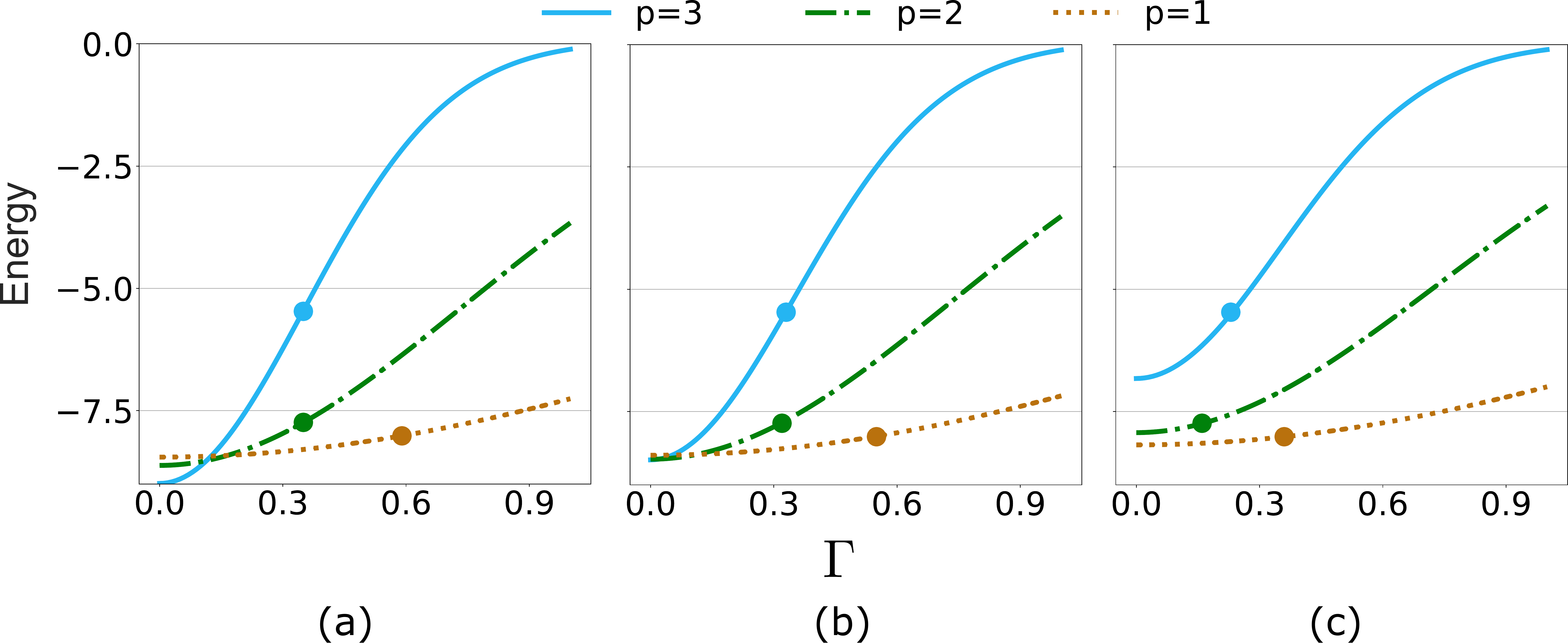}
 \caption{Results from the simulation with noise for the $p=1,2,3$ QAOA protocols for preparation of the critical ground state of the TFIM. (a)$\lambda=0$ (no depolarizing noise); (b)$\lambda=0.1$; (c)$\lambda=0.22$. Each curve is averaged over 1000 samples. The circles in the figure show with a given set of $p$ and $\lambda$, what $\gamma$ value does an experimental result predict. With $\lambda=0.22$ being the point at which we can minimize the variance of the predicted $\Gamma$. Note that (a) shows a threshold at $\Gamma=0.13$ below which higher $p$ give better results.} \label{fig:noise}
 \end{figure*}

For each number of layers $p$, we run the protocol with optimal angles on the trapped ion system and again measure two body correlation functions for Pauli Z and X operators (Fig. \ref{fig:critical_state}(a)(b)). The theoretical and experimental values agree well for the $p=1$ protocol, but deviate for $p=2,3$, as errors accumulate in the deeper circuit.  The energy calculated from correlation functions shows similar accumulation of errors. In experiment, with the p=3 protocol, the generated states can reach energy as low as $-5.46 \pm-0.097$. The lowest energy a p=2 protocol brings a state to is $-7.74\pm0.095$. The lowest energy a p=1 protocol brings a state to is  $-8.02\pm0.043$. In the simulation, the corresponding numbers are -8.98 for p=3, -8.62 for p=2 and -8.44 for p=1. Fig.\ref{fig:noise} (a) provides a visual comparison.
%Hence, there is a tradeoff between using deeper circuits (higher $p$ protocols) to increase fidelity with the target state and the increase in errors associated with deeper circuits; this leads to an error threshold below which an experiment implementing QAOA-type protocols can benefit from higher $p$.   
%An interesting outcome of this experiment is
We find that the QAOA protocol with the least number of steps turns out to be the most successful, producing the state with the lowest energy, though theoretically it should be the worst. This reflects the level of noise in the experimental system which we discuss in the final section of this paper.
%fact that quantum computers/simulators are still in the Noisy Intermediate Scale Quantum era, and that the efficacy of quantum computing algorithms needs to be judged from both a theoretical standpoint and in terms of performance on actual devices. 

%\section{Hybrid Quantum-Classical Approach}

\subsection{Full Hybrid Optimization: Preparation of Ground State of TFIM}

Determining the optimal angles using classical simulation is feasible for current system sizes. For larger systems and higher $p$, however, one would need extrapolation based on patterns in the control parameters of QAOA protocols \cite{zhou2018quantum, brandao2018fixed}.  Therefore, a hybrid approach which involves a feedback loop between a quantum simulator and a classical computer has to be employed. As depicted in Fig. 1, one first carries out the unitary circuit for a given set of parameters, measures the energy cost function, and then uses classical optimization to vary the parameters to lower the cost function until convergence is reached.  One benefit of this hybrid scheme is that systematic errors from the quantum device are reduced. 

We implement the full QAOA hybrid algorithm using standard gradient descent as the classical optimization strategy. To obtain an estimate of the partial derivatives, we change each parameter separately by a small amount and measure the corresponding energy difference. We then take a small(proportional to the gradient, with coefficient adjusted according to simulation) step along the gradient with all parameters. We target the critical TFIM ground state for $p=1$ starting from a random set of initial parameters. Results are shown in figure \ref{fig:critical_state}(c). The optimization converges to a set of parameters that is different from the simulated result, but the measured energy matches the theoretical prediction for $p=1$. 

To examine whether significant systematic errors play a role for deeper circuits in our experiment, we implement the hybrid optimization for $p=2$. This time, we initialize the process with the optimum values obtained from numerical simulation. A drop in the cost function would indicate that systematic errors shift the system away from the optimal state. The results in figure \ref{fig:critical_state}(d) show that this is not the case in our system. 

%\begin{figure}[htbp]
%\centering
% \includegraphics[width=\textwidth]{critical_state_optimization.pdf}
% \caption{Hybrid quantum-classical optimization with QAOA protocols: (a) correspond to the optimization of a $p=2$ protocol, with control parameters $(\vec{\gamma},\vec{\beta})$ initialized according to the optimial set in the simulation. (b) correspond to the opimization of a $p=1$ protocol, with control parameters $(\vec{\gamma},\vec{\beta})$ initialized randomly. In the upper part of (a) and (b), the curve correspond the the energy at each iteration. The dots stands for the measurements taken at each iteration in order to calculate the gradient. The lower part of (a) and (b) show how control parameters  $(\vec{\gamma},\vec{\beta})$ are changed by the optimizer at each iteration.} \label{fig:optimization}
% \end{figure}

\subsection{Error simulation}

We simulate the QAOA protocol in the presence of noise for different numbers of layers $p$, analyzing the trade-off between theoretical and experimental errors.
The two-qubit XX gates are the main source of error in the experiment, likely limited by laser beam intensity fluctuation $\delta I$ on the trapped ion qubits. Because the angle of the XX gate evolution depends on the square of the laser intensity $I$, the fractional error in the XX gate angle is $\Gamma = 2\delta I/I$.
%The intensity, $I$ of the laser beam used to implement them may fluctuate due to power or beam pointing noise, leading to an under- or over-rotation in the angle. The target angle $\theta_0$ is related by a constant $\alpha$ to $I$ which is modeled to fluctuate by an amount $\delta$. Then the error $\Delta$ in the angle can be estimated as:
%\begin{eqnarray}
%    \theta_0&=&I^2 \alpha^2\nonumber\\
%    \theta_0+\Delta&=&(I+\delta)^2\alpha^2\approx (I^2+2\delta I)\alpha^2\nonumber\\
  %  \Delta&=&\Gamma \theta_0,
%\end{eqnarray}
%where $\Gamma=2\delta/I$ sets the magnitude of noise. 
We model this error with a Monte Carlo simulation by setting the angle of the two qubit gate to be $\theta=\theta_0(1+\Gamma r)$, where $\theta_0$ is the nominal gate angle, $r$ is a Gaussian-distributed random number with mean 0 and standard deviation 1, and we average over 1000 samples. Fig. \ref{fig:noise}(a) shows the results for the variation of the measured energy versus $\Gamma$. The 3 points marked in the figure indicate the experimentally measured values for the $p=1,2,3$ protocols. The value of the noise parameter $\Gamma$ inferred from this error model is consistent between $p=2$ and $p=3$. 
%At $p=1$, the curve is relatively flat, and the larger value of $\Gamma$ predicted for this experiment may be due to the error being dominated by another source not proportional to the number of gates, such as the readout.

As seen in Fig. \ref{fig:noise}(a), for $\Gamma\lesssim 0.13$, the higher-depth circuit produces a better outcome, and for higher levels of $\Gamma$, the lower depth circuit is preferable. This implies a type of threshold noise behavior, where the optimization protocol converges to near-optimal solutions as long as the noise is below a critical value.

Generically, we also expect the two-qubit gates to include some depolarizing error on the qubits involved in the gate. This error-channel can be simulated by averaging over rotations $\phi$ around a random axis before every XX gate  (see Appendix for details). We parameterize the rotations by letting $\phi$ be a random variable with distribution $P(\phi)$ that is Gaussian with mean 0 and standard-deviation $\lambda$. The variance in $\gamma$ is calculated for several values of $\lambda$ between $0.1$ and $0.3$. Fig. \ref{fig:noise} (b) and (c) show results for different values of $\lambda$, with $\lambda=0.22$ being the point at which we can minimize the variance of the predicted $\Gamma$. Note that the threshold described above is only present for sufficiently low $\lambda$.

The predicted values of $\Gamma$ and $\lambda$ are in line with experimental accuracy. 
Considering both experimental and theoretical errors for particular algorithms is essential %in the NISQ era 
as quantum computers and simulators scale up.

\section{Outlook}

Our protocols for generating novel thermal states of qubits leverage the recent advent of variational approaches, in particular QAOA, and serve as the first step of several interesting directions. Even without the full hybrid quantum-classical scheme, our theoretical and experimental methods enable the exploration of very interesting physics.  On one hand, the duality between a wormhole and a critical TFD can be taken one major step further: the traversal of the wormhole corresponds to  performing simple operations on the TFD state \cite{traversable1, traversable2}.  In experiment, this traversal could be confirmed by verifying teleportation between the two sides of the TFD.  In a different vein, our critical TFIM ground state preparation paves the way for extracting universal aspects of quantum criticality, such as the central charge of a conformal field theory, from experiments.  Additionally, one could use the TFD protocol to probe the quantum critical fan at finite temperature \cite{subir}.

Our hybrid approach for creating pure (T=0) states of the TFIM system also applies to thermal state preparation. In that case the cost function to be measured is the free energy on system A: $F_A=E_A-TS_A$, where $E_A=\Tr(\rho_A H_A)$ and $S_A=-\Tr(\rho_A \log \rho_A)$ are the energy and entanglement entropy between $A$ and $B$.  Estimating the latter would involve extrapolating from several Renyi entropy measurements, which requires either several copies of the system  \cite{johri_renyi, linke_renyi} or randomized measurements on one copy \cite{Brydges260}. In the longer term, the hybrid approach for both quantum pure and thermal state preparation may enable one to probe many-body physics on system sizes beyond the reach of classical computers and thus shed light on the full (finite temperature) phase diagram of intractable models.  

On a practical level, our hybrid quantum-classical experiment and noise analysis suggest an error threshold that near-term devices must overcome to unlock the full potential of variational approaches. 

% On a practical level, once systematic errors are eliminated, as indicated by our hybrid quantum-classical study, our analysis shows that while variational algorithms may provide new ways to access many-body physics, they are not a trick to circumvent the problem of noisy gates. Our noise analysis indicating a threshold in the error of the two-qubit gates means that simply putting noisy gates together to form a variational circuit is a fruitless enterprise. This is so especially in light of QAOA proofs that indicate that quantum advantage will require going to higher $p$ especially for larger system sizes. It also brings into question the scalability of recent popular approaches such as [IBM paper] on hardware that is noisy enough to be above such a threshold. 

\section{Acknowledgements}
We would like to thank A. Seif for helpful discussions. This work is supported in part by the ARO through the IARPA LogiQ
program, the AFOSR MURIs on Quantum Measurement/Verification and Quantum Interactive Protocols, the ARO MURI on Modular Quantum Circuits, the DoE BES QIS Program, the DOE HEP QuantISED Program, the DoE ASCR Quantum Testbed program, and the NSF Physics Frontier Center at JQI.   Research at Perimeter Institute is supported
by the Government of Canada through Industry Canada
and by the Province of Ontario through the Ministry of
Research and Innovation. C.H.A. acknowledges financial support from CONACYT doctoral grant No. 455378.

\section{Appendix}

\subsection{Simulating depolarizing noise}
Here, we discuss our simulation of depolarizing noise. This noise is due to residual entanglement of the qubits with the motional modes at the end of a gate operation. It error-channel can be simulated by averaging over rotations around a random axis after every XX gate.
To see this, first consider the effect of a depolarizing channel on the density matrix for a single qubit:
\begin{align}\label{eq:depol}
    \rho\xrightarrow{\text{depol}}\bigg(1-\frac{3p}{4}\bigg)\rho +\frac{p}{4}(\sigma_{X}\rho \sigma_{X}+\sigma_{Y}\rho \sigma_{Y} +\sigma_{Z}\rho \sigma_{Z}),
\end{align}
where $\sigma_{X/Y/Z}$ are the Pauli matrices.

Instead, rotating by an angle $\phi$ around an axis $\hat{n}$ would give:

\begin{widetext}
\begin{eqnarray}
    \rho &\xrightarrow{}& \exp(i\frac{\phi}{2}\hat{n}.\vec{\sigma}) \rho \exp(-i\frac{\phi}{2}\hat{n}.\vec{\sigma})\nonumber\\
    &=&\bigg(\cos(\frac{\phi}{2})I +i\sin(\frac{\phi}{2})\hat{n}.\vec{\sigma}\bigg) \rho \bigg(\cos(\frac{\phi}{2})I -i\sin(\frac{\phi}{2})\hat{n}.\vec{\sigma}\bigg)
\end{eqnarray}
\end{widetext}

Here $\hat{n}.\vec{\sigma}=n_X\sigma_X+n_Y\sigma_Y+n_Z\sigma_Z$. Let $\phi$ be a random variable with distribution $P(\phi)$ that is Gaussian with mean 0 and standard-deviation $\lambda$. Averaging over samples with different values of $\phi$ and $\hat{n}$ is equivalent to integrating $\int d\hat{n}\int_{-\pi}^{\pi} P(\phi)\rho d\phi$, where $P(\phi)$ is the distribution over $\phi$. On integration, all the terms containing one $\sin(\phi)$ term will disappear since they are odd functions of $\phi$. On integrating over $n_{X/Y/Z}$, the only terms that remain are the ones that contained $n^2_{x/y/z}$ and so are of the form $\sigma_X\rho\sigma_X$. So finally this procedure returns the single qubit depolarizing channel in Eq. \ref{eq:depol} with $p$ a function of $\lambda$. This treatment can be straightforwardly extended to a depolarizing channel on two qubits by averaging over rotations around random axes on both qubits after every XX gate.

%\begin{widetext}

\begin{figure*}[t]
\centering
\includegraphics[width=0.66\textwidth]{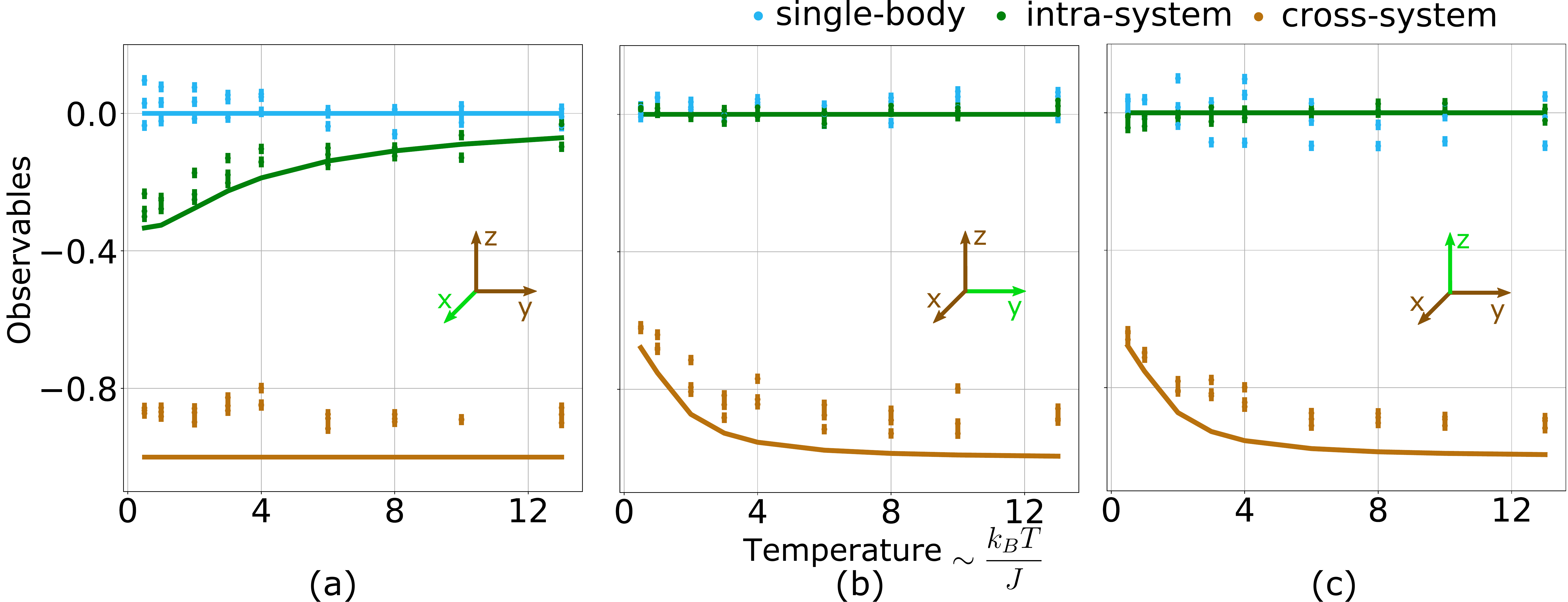}
 \caption{Classical Ising TFD states on a 6-qubit system.  The measured one body and two body correlation functions for different target temperatures are plotted against the theoretical expectations (dotted lines) for type (a) Pauli-X  (b) Pauli-Y and (c) Pauli-Z. Intra-system correlators are two body correlators in the subsystem-A: $<\sigma_{1,A}\sigma_{2,A}>$, $<\sigma_{1,A}\sigma_{3,A}>$, and $<\sigma_{2,A}\sigma_{3,A}>$.Cross-system correlators are $<\sigma_{1,A}\sigma_{1,B}>$, $<\sigma_{2,A}\sigma_{2,B}>$, and $<\sigma_{3,A}\sigma_{3,B}>$. Note the experimental data points in the figure have errorbar accounting to statistical errors. Due to the number of samples taken, the error bars are the same size or smaller than the symbols used.} \label{fig:TFD}
 \end{figure*}
 
% \end{widetext}

\subsection{Experimental Details}

The system is based on a chain of ${}^{171}\textrm{Yb}^+$ ions held in an RF Paul trap \cite{debnath2016demonstration}. Each ion provides one physical qubit in the form of a pair of states in the hyperfine-split ${}^{2}\textrm{S}_{1/2}$ ground level with an energy difference of 12.642821 GHz, which is insensitive to magnetic field fluctuations to first order. The qubits are initialized to $\ket{0}$ by optical pumping and read out by state-dependent fluorescence detection \cite{Olmschenk07}. Gates are realized by a pair of Raman beams derived from a single 355-nm mode-locked laser. These optical controllers consist of a global beam that illuminates the entire chain and an array of individual addressing beams. Single-qubit rotations are realized by driving resonant Rabi rotations of defined phase, duration, and amplitude. Two-qubit gates are achieved by illuminating two selected ions with beat-note frequencies near the motional sidebands creating an effective Ising spin-spin interaction via transient entanglement between the two qubits and the motion in the trap \cite{Molmer99,Solano99,Milburn00}. Our scheme involves multiple modes of motion, which are disentangled from the qubits at the end of an two-qubit gate operations via an amplitude modulation scheme\cite{choi2014optimal}. Typical single- and two-qubit gate fidelities are $99.5(2)\%$ and $98-99\%$, respectively. The latter is  limited by residual entanglement of the qubit states and the motional state of the ions due to intensity noise, and motional heating. Rotations around the z-axis are achieved by phase advances on the classical control signals.

The initialization of Bell-pairs between qubits $i$ and $j$ is implemented through the following sequence of gates:
\begin{equation}\label{eq:Bell_Pair}
  \frac{1}{\sqrt{2}}(\ket{0}_i\ket{1}_j-\ket{1}_i\ket{0}_j)=RZ_i(\frac{\pi}{2})RX_i(-\pi)XX_{i,j}\ket{0}_i\ket{1}_j
\end{equation}
Here $RZ$ and $RX$ stand for single qubit rotation gates about $Z$ and $X$ axis, respectively, and $XX$ for entangling Ising gates.

The unitaries required for the QAOA protocols (Eq. \ref{eq:QAOA_process}), $\exp(iH_{ABX}\alpha_1)$ and $\exp(iH_{XX}\gamma_2)$ are directly implemented as $XX$ gates. $\exp(iH_{ABZ}\alpha_2)$ is implemented by converting $XX$ gates into $ZZ$ gates through single qubit x-rotations. 

\subsection{Thermofield Double State of Classical Ising Model}

For the one-dimensional classical Ising model with periodic boundary conditions, $H_A=\sum_{i=1}^L X_i X_{i+1}$, the protocol can be simplified due to the extensive number of conserved quantities.  As demonstrated in \cite{wu2018variational}, the simple sequence
\begin{eqnarray}
| \psi(\vec{\alpha},\vec{\gamma})\rangle_p = \prod_{i=1}^{p} e^{i\alpha_i H_{ABZ}} e^{i\gamma_i H_A} \ket{\psi_0}, 
\end{eqnarray}
where $H_{ABZ}\equiv \sum_i Z_{i,A} Z_{i,B}$, is sufficient to produce the classical Ising TFD state on a system of $L$ (by 2) qubits given $p=L/2$ iterations.

For a system of in total 6 qubits ($L=3$), the above protocol only requires one application of $H_A$ and $H'_{AB}$ ($p=1$) to perfectly prepare the target TFD at any temperature.  For each target temperature, the optimal timesteps $\alpha_1, \gamma_1$ are obtained (on a classical computer) by maximizing the fidelity cost function.  

We implement these protocols for TFDs at various temperatures on the trapped ion system.
To verify the TFD preparation, we measure the two-point correlation functions $\langle X_{i,A}X_{i,B}\rangle,\langle Y_{i,A}Y_{i,B}\rangle,\langle Z_{i,A}Z_{i,B}\rangle$ between the A and B subsystems, the intra-system correlators $\langle X_{i,A}X_{j,A}\rangle$, and single-body observables, see Fig.\ref{fig:TFD}.

\subsection{Symmetry Based Error Mitigation}

\begin{figure*}[t]
\centering
\includegraphics[width=0.66\textwidth]{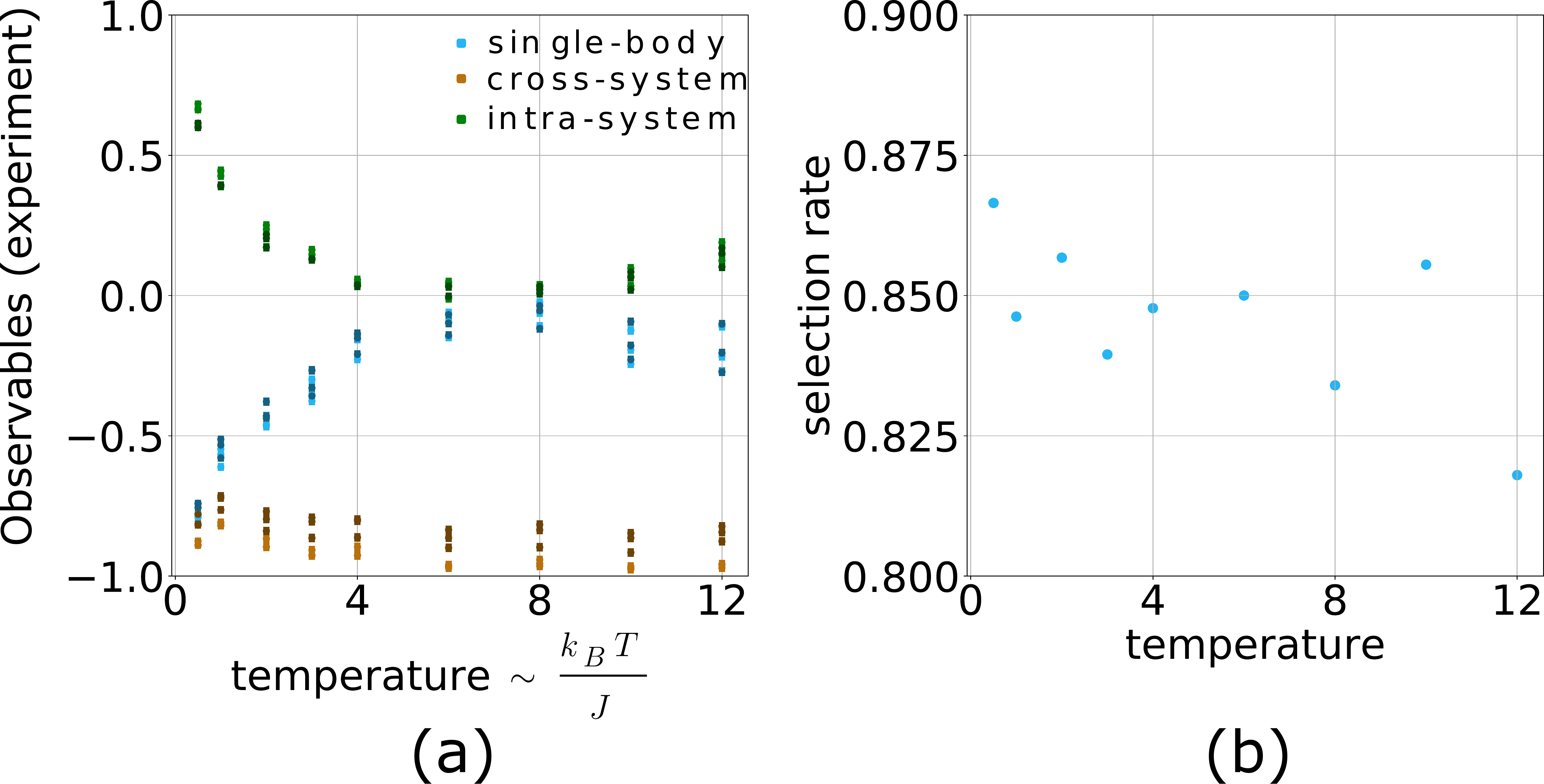}
 \caption{(a): Comparison between results with and without symmetry based error mitigation. The dots with same hue correspond to the same type of measurement, while dots with brighter color correspond to data corrected with error mitigation.(b): The fraction of data points kept after the symmetry based post-selection.} \label{fig:Error_Mitigation}
 \end{figure*}
 
 The transverse field Ising chain defined as $H=\sum_i (X_i X_{i+1} + g Z_i)$ has a $Z_2$ symmetry, i.e. the Hamiltonian commutes with the operator $\prod_i Z_i$. The TFD state $|\Psi\rangle=\frac{1}{Z(\beta)}\sum_n exp(-\beta E_n/2)|n\rangle_A|n'\rangle_B$ is a superposition of states in which subsystem B has a time-reversed copy of the eigenstate of $H$ in A. Therefore, $Z_{1A} Z_{2A} Z_{3A}=-Z_{1B} Z_{2B} Z_{3B}$, and any measurement that does not satisfy this should be discarded.
 
 This symmetry based error mitigation is applied to measurements in the $Z$ direction. Fig. \ref{fig:Error_Mitigation} (a)  visualizes the difference between the corrected and uncorrected data. Notable improvement can be seen in the cross-system correlators. Fig.\ref{fig:Error_Mitigation} (b) visualizes the selection rate (the proportion of data kept) at each temperature. It can be observed from Fig.\ref{fig:Error_Mitigation} (b) that the selection rate drops as the temperature rise. This agrees with the trend observed in Fig.\ref{fig:TFD_TFIM} that the error is larger at high temperature.

%\bibliography{reference}

\end{document}